\documentclass[a4paper,twoside,notoc,11pt]{JHEP3}

\usepackage{epsfig,multicol} \usepackage{delarray,amsmath,bbm}

\newcommand\fverb{\setbox\pippobox=\hbox\bgroup\verb} \newcommand\fverbdo{\egroup\medskip\noindent% \fbox{\unhbox\pippobox}\
} \newcommand\fverbit{\egroup\item[\fbox{\unhbox\pippobox}]}
\newbox\pippobox

\newcommand{\nn}{\nonumber} \newcommand{\beq} {\begin{equation}}
\newcommand{\eeq} {\end{equation}} \newcommand{\beqa}
{\begin{eqnarray}} \newcommand{\eeqa} {\end{eqnarray}}

\newcommand{\ie}{{\it i.e.}}  
  \newcommand{\etal}{{\it et al.}}

  \newcommand{\morder}[1]{{\cal O}\left(#1 \right)}
\newcommand{\eq}[1]{(\ref{#1})}

  \newcommand{\halft}{{\textstyle \frac{1}{2}}}
 \newcommand{\gsim}{\gtrsim}
 \newcommand{\im}{{\,\rm Im \,}}
\newcommand{\re}{{\,\rm Re \,}}

\title{\center{Retardation Effect for Collisional Energy Loss \\ of
Hard Partons Produced in a QGP}}

\author{St\'ephane Peign\'e\footnote{On leave of absence from LAPTH,
CNRS, UMR 5108, Universit\'e de Savoie, B.P. 110, F-74941
Annecy-le-Vieux Cedex, France}, Pol-Bernard Gossiaux, Thierry
Gousset\\ SUBATECH, UMR 6457, Universit\'e de Nantes \\ Ecole des
Mines de Nantes, IN2P3/CNRS. \\ 4 rue Alfred Kastler, 44307 Nantes
cedex 3, France \\ E-mail: \email{lastname@subatech.in2p3.fr}}

\preprint{SUBATECH 2005/004 \\ LAPTH-1110/05}

\abstract{We study the collisional energy loss suffered by an
energetic parton travelling the distance $L$ in a high temperature
quark-gluon plasma and {\it initially produced in the medium}. We
find that the medium-induced collisional loss $-\Delta E(L)$ is
strongly suppressed compared to previous estimates which assumed the
collisional energy loss rate $-dE/dx$ to be constant. 
The large $L$ linear asymptotic behaviour of $-\Delta E(L)$ sets in only after a
quite large retardation time.  The suppression of $-\Delta E(L)$ is 
partly due to the fact that gluon bremsstrahlung arising from the
initial acceleration of the energetic parton is reduced in the medium
compared to vacuum.  The latter radiation spectrum is sensitive to the
plasmon modes of the quark-gluon plasma and has a rich angular
structure.}


\begin{document}

\section{Introduction}

Jet quenching has long ago been suggested as a possible signal for the
quark-gluon plasma (QGP) \cite{bj}, and this has triggered a
considerable activity both on the experimental and theoretical sides.
The theoretical determination of parton energy loss has been the
subject of many studies, and numerous observables sensitive to jet
quenching are currently analyzed in the Relativistic Heavy Ion
Collider (RHIC) experiments.

In a (static) thermal or dense medium of very large size $L$, a parton
of high (but finite) energy $E$ undergoes a radiative energy loss
$\Delta E_{rad}$ which increases as a power of the energy ($\Delta
E_{rad} \propto L E$ in the Bethe-Heitler limit and $\Delta E_{rad}
\propto L \sqrt{E}$ when the non-abelian LPM effect is at work
\cite{bdmps94}), whereas its collisional energy loss $\Delta E_{coll}$
behaves at most lo\-ga\-ri\-thmi\-cally\footnote{The $\log{E}$ factor
arises when $\Delta E_{coll}$ is calculated in the fixed coupling
approximation \cite{bj,TG,BT}, but is expected to be absent in the
case of a running coupling \cite{peshierdok}. This point is briefly
discussed in section 3.2.}, $\Delta E_{coll} \propto L \log{E}$
\cite{bj,TG,BT}. For a large size medium we thus expect $\Delta
E_{rad}$ to be the dominant source of energy loss for energetic
partons.

In the opposite limit, namely for a parton of asymptotic energy $E\to
\infty$ crossing a medium of {\it finite} size $L < L_{cr}$ (but still
thick enough, $L/\lambda \gg 1$, where $\lambda$ is the parton mean
free path), $\Delta E_{rad}$ becomes independent of $E$, $\Delta
E_{rad} \propto E^0 L^2$ \cite{bdmps96,Zakharov}. The finite size
limit $L < L_{cr}$ should be relevant in practice since a simple
numerical estimate within the model of Ref.~\cite{bdmps96} gives
$L_{cr} = \sqrt{\lambda E /m_D^2} \simeq 7 \ {\rm fm}$ for $E = 10 \
{\rm GeV}$, where $m_D^{-1}$ is the Debye screening length in the
medium. It was also shown \cite{GLV} that finite opacity ($L/\lambda
\gsim 1$) and finite $E$ effects lead to a substantial suppression
(together with some energy dependence) of $\Delta E_{rad}$, when
compared to the asymptotic $E$-independent result. We can thus expect
$\Delta E_{coll}$ to compete with $\Delta E_{rad}$ in the case of a
finite size medium.  Indeed, recent studies
\cite{MustafaThoma,dars,Mustafa} suggest that for `jets' of energy on
the order of $10 \,\rm{GeV}$ such as those measured in $AA$ collisions
at RHIC, the collisional energy loss might be comparable to the
radiative one, both for light \cite{dars} and heavy \cite{Mustafa}
partons.

Thus an accurate determination of $\Delta E_{coll}$ for large $E$ and
finite $L$ is needed in order to interpret the suppression of
inclusive large $p_T$ hadron production at moderate energies observed
at RHIC in $AA$ collisions \cite{phenix,star}.  In previous studies of
parton collisional energy loss in a QGP, the loss\footnote{From now on
all energy losses will be implicitly collisional, $-\Delta E \equiv
-\Delta E_{coll}$.} $-\Delta E(L)$ suffered by the hard parton
travelling the distance $L$ in the plasma was assumed to be given by
$(-dE/dx)_{\infty}$ times $L$, with $(-dE/dx)_{\infty}$ the rate of
energy loss per unit distance occurring in a {\it stationary} regime,
\ie, long after the energetic parton has been produced. This is
certainly a valid approximation when the medium size becomes very
large, and formulas for $(-dE/dx)_{\infty}$ of a heavy quark of mass
$M \gg T$ travelling in a QGP of high temperature $T$ have been
obtained \cite{TG,BT}.

In the present paper we argue that such an approximation to $-\Delta
E(L)$ is incorrect in the case of an energetic parton {\it produced at
an initial time $t = 0$} inside a thermal or dense medium of moderate
size $L$, which is the relevant situation when discussing jet or
hadron quenching at large $p_T$. As in previous studies of collisional
energy loss \cite{TG,BT}, we work in the theoretical high temperature
small coupling limit $g \ll 1$ for the QGP (and in the hard thermal
loop - HTL - resummation framework \cite{pisarski,BI}), and assume a
fixed coupling $\alpha_s$. We also consider a static (non-expanding)
QGP in thermal equilibrium.
As argued in section 3.2, we expect the retardation of the stationary
regime found in this paper to be qualitatively unchanged in the case
of a running coupling. However, a rigorous treatment taking into
account the running of $\alpha_s$ would be needed in order to obtain a
reliable quantitative estimate of $-\Delta E(L)$.  Recalling moreover
that $g(m_D) \gsim 1$ in realistic phenomenological applications, we
stress that our results should be considered on a {\it qualitative}
level only.

For a fast parton prepared at $t=-\infty$ and travelling in an
infinite medium, the collisional energy loss can be understood as
follows. The proper (chromo-)electric field of the parton polarizes
the medium, which creates an effective (medium-induced) electric field
around the parton, responsible for its slowing down. If the energetic
parton is produced at $t = 0$ (via some process involving a hard scale
$\sim p_T$), we may expect the rate of collisional energy loss
$-dE/dx$ to be reduced during the time the parton proper field is
created, before reaching the asymptotic value $(-dE/dx)_{\infty}$. A
result suggestive of such a {\it retardation effect} is qualitatively
what we find in the following. A fast parton produced initially in the
medium needs to travel some distance before losing energy at the
highest rate. Our main conclusion is that {\it collisional} losses
used in the analysis of nuclear modification factors $R_{AA}$ at large
$p_T$ (for jet or hadron production) have been systematically
overestimated due to the neglect of this effect.

In section 2 we present our model for the induced collisional energy
loss $-\Delta E(L)$ of a parton produced initially in a QGP.  In
section 3 we give our numerical results for the $L$-dependence of
$-\Delta E(L)$, which display a strong attenuation of the energy loss
when compared to previous estimates, as well as a large 
retardation time $t_{\rm ret}$ of the stationary regime. 
We give a simple interpretation of the largeness of $t_{\rm ret}$.  
We also discuss the domain of validity of our analysis, and 
how the running of $\alpha_s$ could affect 
$-\Delta E(L)$. Finally, in section 4, we show that within our (standard)
definition of {\it collisional} energy loss, $-\Delta E(L)$ gets a
contribution from induced gluon {\it radiation} which arises from the
sudden acceleration of the parton at $t=0$.  The in-medium
bremsstrahlung spectrum due to charge acceleration is indeed not the
same as in vacuum since it is sensitive to the longitudinal and
transverse plasmon modes of the QGP. The angular spectrum presents a
diffraction pattern depending on the plasma size $L$. The {\it
induced} radiated energy is negative, and thus contributes (but only partly) to
the suppression of $-\Delta E(L)$ and to the retardation effect
discussed in section 3. We conclude in section 5.

\section{Model for collisional energy loss of a parton produced at
  $t=0$ in a QGP}

We derive in this section the master equation \eq{mastereq} for the
collisional energy loss $-\Delta E(L)$ of an energetic (and massive)
parton of velocity $v$, initially produced in a quark-gluon plasma,
and travelling the distance $L$ in the medium.  We first present our
model for the classical partonic current density $j^{\mu
\,a}=(\rho^a,\vec{j}^a)$, and then give, in the abelian approximation,
the expression of the (chromo-)electric field induced by this current
density.

\subsection{Model for partonic current}
\label{sec2.1}

In previous studies of parton collisional energy loss in a QGP
\cite{TG,BT}, the classical color charge has been assumed to be
produced at $t=-\infty$. In this case the current 4-vector in
coordinate space reads: \beq
\label{jinf}
j^{\mu\,a}_{\infty}(t,\vec{x}) = q^a V^{\mu} \delta^3(\vec{x} -
\vec{v}\,t) \ , \eeq where $V^{\mu}$ denotes the parton 4-velocity,
$V= (1,\vec{v})$.  The color index $a$ is carried by the parton color
charge $q^a$ defined by $q^a q^a = C_R \alpha_s$, where $C_R$ is the
Casimir invariant of the color representation $R$ the parton belongs
to ($C_R=C_F=4/3$ for a quark and $C_R=C_A=3$ for a gluon). In
4-momentum space $K=(\omega, \vec{k})$ the current \eq{jinf} becomes
\beq
\label{jinfmom}
j^{\mu\,a}_{\infty}(K) = 2\pi q^a V^{\mu} \delta(K.V) \ ,
\eeq
which trivially satisfies current conservation, $K.j_{\infty}=0$.

When considering the quenching of large $p_T$ jets or hadrons in
relativistic heavy ion collisions, the large $p_T$ parent parton is
rather created at $t=0$ in the underlying hard partonic subprocess. 
Its associated classical `current' is thus, instead of \eq{jinf}, 
\beq
\label{j}
j_0^{\mu\,a}(t,\vec{x}) = q^a V^{\mu} \delta^3(\vec{x} - \vec{v}\,t)
\, \theta(t) \ ,
\eeq
giving in momentum space: \beq
\label{jmom}
j_0^{\mu\,a}(K) = i q^a \frac{V^{\mu}}{K.V + i\eta} = q^a V^{\mu}
\left[i {\rm P}\left(\frac{1}{K.V}\right) + \pi \delta(K.V) \right] \
.  \eeq However, the `current' \eq{jmom} is not conserved, $K.j_0 \neq
0$. Color charge conservation requires the hard parton to be produced
in conjunction with at least another parton in the partonic
subprocess.  As a generic {\it conserved} partonic current, we will
consider the simple case of a (color singlet) dipole produced at
$t=0$, consisting of two partons (with the same color charge $q^a$) of
4-velocities $V_1 = (1, \vec{v}_1)$ and $V_2 = (1, \vec{v}_2)$: \beq
\label{genericcurrent}
j^{\mu\,a} = (\rho^a, \vec{j}^a) = i q^a \,
\left(\frac{V_1^{\mu}}{K.V_1+i\eta} - \frac{V_2^{\mu}}{K.V_2+i\eta}
\right) \ .  \eeq In the following we will consider the second parton
to be a static heavy quark, \ie\ $\vec{v}_2= \vec{0}$, in which case
the space component of the current is simply given by the first term
of \eq{genericcurrent}. The choice \eq{genericcurrent} instead of
\eq{jmom} is however crucial.  Current conservation indeed constrains
the form of the electric field \eq{linappr}, and allows to treat
unambiguously the potential singularity at $\omega=0$ (see
\eq{linappr} and \eq{efield}), as we briefly explain below. We expect
the main results of our study not to depend on the simplifying
assumption $\vec{v}_2= \vec{0}$.

\subsection{Induced electric field} 
\label{sec2.2}

Following \cite{TG} (see also \cite{Ichimaru}), in the abelian
approximation and within linear response theory the Maxwell equations
can be solved in 4-momentum space $K=(\omega, \vec{k})$, yielding the
total (chromo-)electric field $\vec{E}^a$ in terms of the classical
vector current density $\vec{j}^a$: \beq
\label{linappr}
\epsilon_L \vec{E}_L^a + (\epsilon_T - k^2/\omega^2) \vec{E}_T^a =
\frac{4\pi}{i\omega} (\vec{j}_L^a + \vec{j}_T^a) \ .  \eeq The
longitudinal and transverse components are given by $\vec{j}_{L} =
(\vec{j}.\vec{k}/k^2)\vec{k}$ (we denote $k=|\vec{k}|$) and
$\vec{j}_{T} = \vec{j}-\vec{j}_{L}$. In the linear response
approximation the longitudinal and transverse dielectric functions of
the plasma $\epsilon_L$ and $\epsilon_T$ are not affected by the
external current. We consider a high temperature QGP, for which
$\epsilon_{L,T}$ have been obtained in \cite{klimov,weldon} and later
rederived in the gauge-invariant HTL resummation framework
\cite{pisarski,BI}.

In the following we will have to deal with the $1/\omega$ potential
singularity appearing in \eq{linappr}, and specifically affecting the
longitudinal part of the electric field. The latter actually arises
from Coulomb's law $k \, \epsilon_L E_L \propto \rho$, where $\rho$ is
the charge density, by using the equation for current conservation
$\rho = k j_L / \omega$. More precisely, our conserved current
\eq{genericcurrent} satisfies $\rho = k j_L/(\omega +i\eta)$, for any
$\vec{v}_1$ and $\vec{v}_2$ (including $\vec{v}_2=\vec{0}$). This
shows that the potential $1/\omega$ singularity appearing in
\eq{linappr} should be regularized with the {\it retarded}
prescription, $\omega \rightarrow \omega +i\eta$.

We obtain from \eq{linappr} the {\it medium-induced} electric field
$\vec{\cal{E}}^a$, \beq
\label{efield}
\vec{\cal{E}}^a(t, \vec{x}) = \int_{-\infty}^{\infty}
\frac{d\omega}{\omega} \int \frac{d^3\vec{k}}{4\pi^3 i} \, e^{-i
(\omega t - \vec{k}.\vec{x})} \left[ \frac{\vec{j}_{L}^a}{\epsilon_L}
+ \frac{\vec{j}_{T}^a}{\epsilon_T - k^2/\omega^2} \right]_{\rm ind} \
, \eeq where $\vec{j}^a$ is given by \eq{genericcurrent} (with
$\vec{v}_2= \vec{0}$), or equivalently by \eq{jmom} (with
$\vec{v}=\vec{v}_1$), and the $1/\omega$ singularity should be treated
with the retarded prescription.  In \eq{efield} the subscript denotes
the implicit subtraction of the vacuum contribution (corresponding to
$\epsilon_L=\epsilon_T=1$).  Since the dielectric functions (and the
external current) are real quantities in coordinate space, implying in
momentum space $\epsilon_{L,T}(-K)= \epsilon_{L,T}(K)^*$ (and a
similar relation for the current), the expression \eq{efield} is
easily checked to be also real.

\subsection{Master equation for parton collisional energy loss}
\label{sec2.3}

During its travel in the plasma between $t=0$ and $t=L/v$, the induced
energy gain $\Delta E$ of the parton of {\it constant} velocity
$\vec{v}_1 = \vec{v}$ equals the work of the electric force on its
trajectory, namely $\Delta E = \vec{v} \cdot \int_0^{L/v} dt\, q^a
\vec{\cal{E}}^a(t, \vec{x}= \vec{v} t)$ or: \beq
\label{eloss}
\Delta E(L) = q^a \vec{v} \cdot \int_{-\infty}^{\infty}
\frac{d\omega}{\omega} \int \frac{d^3\vec{k}}{4\pi^3 i} \int_0^{L/v}
dt \, e^{-i K.V\, t} \left[ \frac{\vec{j}_{L}^a}{\epsilon_L} +
\frac{\vec{j}_{T}^a}{\epsilon_T - k^2/\omega^2} \right]_{\rm ind} \ .
\eeq We stress that the latter expression is valid in the abelian
approximation for the hard parton dynamics and within linear response
theory, implying that $|\Delta E(L)|$ should be small compared to the
initial parton energy $E$, which is also consistent with the
assumption of a constant velocity.

If as in \cite{TG} the current \eq{jinfmom} is used in \eq{eloss}, the
exponential factor equals unity, and the collisional energy loss on
the distance $L$ is uniquely determined by its rate per unit distance
$(-dE/dx)_{\infty}= -\Delta E/L$.  Inserting instead
\eq{genericcurrent} in \eq{eloss}, and performing the time integral,
we obtain the collisional energy loss $-\Delta E(L)$ of a hard parton
produced at $t=0$ and travelling the distance $L$ in the medium, \beq
\label{mastereq}
\frac{-\Delta E(L)}{C_R \alpha_s} = i \int_{-\infty}^{\infty}
\frac{d\omega}{\omega} \int \frac{d^3\vec{k}}{4\pi^3} \, \left[
\frac{\vec{v}_{L}^{\,2}}{\epsilon_L}+
\frac{\vec{v}_{T}^{\,2}}{\epsilon_T - k^2/\omega^2} \right]_{\rm ind}
\left\{ \frac{1-e^{-i K.V \, L/v}}{K.V(K.V+i\eta)} \right\} \ .  \eeq
We note that the factor between the curly brackets in \eq{mastereq}
can be rewritten as \beq
\label{bracket}
\left\{ \pi \, \delta(K.V) L/v + 2\,\frac{\sin^2(K.V
\,L/(2v))}{(K.V)^2} + i\,\frac{\sin(K.V \,L/v)}{K.V} {\rm
P}\left(\frac{1}{K.V}\right) \right\} \ \ .  \eeq Using the following
identities, \beq
\label{delta}
\frac{\sin(u L)}{u} \mathop{\longrightarrow}_{L \to \infty} \pi
\delta(u) \ \ \ \ ; \ \ \ \ \frac{\sin^2(u L)}{L\,u^2}
\mathop{\longrightarrow}_{L \to \infty} \pi \delta(u) \ , \eeq we find
that in the $L \to \infty$ limit, the expression \eq{bracket} is
equivalent to $2\pi \, \delta(K.V) L/v$, and \eq{mastereq} thus
reproduces the result for $(-dE/dx)_{\infty}$ obtained in
Ref.~\cite{TG}.  In the small $L$ limit, however, the second term of
\eq{bracket} is subleading and \eq{bracket} reduces to $(i {\rm
P}(1/K.V) +\pi \delta(K.V))\,L/v$ as can be seen also directly from
\eq{mastereq}, thus leading to a modification of collisional energy
loss at finite $L$.

\section{The retardation effect}

In this section we first express the energy loss \eq{mastereq} in
terms of the discontinuity (on the real axis) of the longitudinal and
transverse thermal gluon propagators. We then present and discuss our
numerical results, which show that the asymptotic large $L$ behaviour
of $-\Delta E(L)$ sets in only after some retardation time.

\subsection{Expression of $-\Delta E(L)$ in terms of thermal gluon
  spectral densities}
\label{sec3.1}

The dielectric functions can be expressed in terms of the longitudinal
and transverse thermal gluon self-energies\footnote{In the following,
we will use the sign conventions and notations of Ref.~\cite{BI}.},
\beq
\label{dielectric}
\epsilon_L = 1+\Pi_L(x)/k^2 \ \ \ \ ; \ \ \ \ \epsilon_T =
1-\Pi_T(x)/\omega^2 \ , \eeq where $x=\omega/k$ and $\Pi_{L,T}$ have
been obtained in the HTL approximation \cite{pisarski,BI}, \beqa
\Pi_L(x) = m_D^2 \left[ 1 - \frac{x}{2} \log{\left( \frac{x+1}{x-1}
\right)} \right] \ \ \ ; \ \ \ \Pi_T(x) = \halft m_D^2 \, x^2 \left[1
- \frac{x^2-1}{2x} \log{\left( \frac{x+1}{x-1} \right)} \right] \
. \nn \\
\label{pilt}
\eeqa The Debye mass denoted as $m_D$ is given by $m_D^2 = 4\pi
\alpha_s T^2 (1+n_f/6)$ (with $n_f=2$ the number of thermally
equilibrated flavours). We also use the longitudinal and transverse
thermal gluon propagators \beq
\label{prop}
\Delta_L(\omega=kx, k) = \frac{-1}{k^2+\Pi_L(x)} \ \ \ \ ;\ \ \ \
\Delta_T(\omega=k x, k) = \frac{-1}{\omega^2-k^2-\Pi_T(x)} \ \ .  \eeq
Using \eq{dielectric} and \eq{prop} the expression \eq{mastereq}
becomes (with $\vec{v}_{L}^{\,2} = v^2 \cos^2{\theta}$,
$\vec{v}_{T}^{\,2} = v^2 \sin^2{\theta}$): \beqa
\label{r1}
\frac{-\Delta E(L)}{C_R \alpha_s} &=& - i v^2 \int
\frac{d^3\vec{k}}{4\pi^3} \int_{-\infty}^{\infty}
\frac{d\omega}{\omega} \, \left[ k^2 \cos^2{\theta} \,
\Delta_L(\omega, k)+ \omega^2 \sin^2{\theta} \, \Delta_T(\omega, k)
\right]_{\rm ind} \nn \\ && \hskip 3cm \times \left\{ \frac{1-e^{-i
(\omega-kv\cos{\theta}) \,
L/v}}{(\omega-kv\cos{\theta})(\omega-kv\cos{\theta}+i\eta)} \right\} \
.  \eeqa

The longitudinal and transverse thermal gluon propagators have
singularities on the real $\omega$-axis, namely branch cuts
(corresponding to Landau damping) in the spacelike $|x|<1$ region, and
poles corresponding to collective excitations of the plasma (plasmons)
in the timelike $|x|>1$ region.  Those singularities must be treated
using the retarded prescription $\omega \to \omega +i\eta$ arising in
the analytical continuation from imaginary to real frequencies in
finite temperature field theory.  As we explained in section
\ref{sec2.2}, the potential $1/\omega$ singularity at $\omega \to 0$
must be also regularized with the retarded prescription. Thus all
singularities on the real axis appearing in \eq{r1} are implicitly
written with the same, retarded prescription.

It is convenient to perform the $\omega$-integral in \eq{r1} using
Cauchy's theorem, by closing the integration contour in the lower
(complex $\omega$) half-plane, as required by the presence of the
exponential factor in the integrand (since $L >0$).  Thus the integral
over the real axis is replaced by the integral over the
(clockwise-going) contours around the singularities which lie just
below the real axis, namely the poles at $\omega = -i\eta$ and $\omega
= kv\cos{\theta}-i\eta$, and the plasmon poles and cuts of the
propagators $\Delta_{L,T}$. We obtain from \eq{r1}: \beqa
\label{r2}
\frac{-\Delta E(L)}{C_R \alpha_s} &=& - i v^2 \int
\frac{d^3\vec{k}}{4\pi^3} \left\{
(-2i\pi) \frac{1}{v^2} (1-e^{ikL\cos{\theta}}) \re{\Delta_L(0, k)}
\phantom{{\rm P}\left( \frac{1}{\omega} \right)} \right. \nn \\
&& + (-2i\pi) \frac{iL k \cos{\theta}}{v^2}
\left[\re{\Delta_L(kv\cos{\theta}, k)}
+v^2 \sin^2{\theta} \re{\Delta_T(kv\cos{\theta}, k)} \right] \nn \\
&& + 2 i \int_{-\infty}^{\infty} d\omega  \, 
\left[ k^2 \cos^2{\theta} \im{\Delta_L(\omega, k)}+ \omega^2
  \sin^2{\theta} \im{\Delta_T(\omega, k)} \right] \nn \\
&& \hskip 2cm \times  \left. {\rm P}\left( \frac{1}{\omega} \right)
     {\rm P}\left( \frac{1}{\omega- kv\cos{\theta}} \right) 
\frac{1-e^{-i (\omega-kv\cos{\theta}) \, L/v}}{\omega-kv\cos{\theta}}
\right\}_{\rm ind} \ .
\eeqa
Due to the fact that $\re{\Delta_{L,T}(\omega, k)}$ is an even
function of $\omega$, the second line of the latter equation vanishes
after angular integration. Recalling \cite{BI} that the spectral
densities (with $s=L$ or $T$) \beq
\label{rholt}
\rho_{s}(\omega, k) \equiv 2 \im{\Delta_{s}(\omega + i\eta, k)} = 2\pi
{\, \rm sgn}(\omega) \,z_s(k) \delta(\omega^2 -\omega_s^2(k)) +
\beta_{s}(\omega, k) \theta(k^2-\omega^2) \eeq vanish at $\omega =0$,
we can rewrite \eq{r2} in the form: \beqa
\label{mastereq2}
\frac{-\Delta E(L)}{C_R \alpha_s} = \int \frac{d^3\vec{k}}{2\pi^2}
\left\{ \frac{1-\cos{(kL\cos{\theta})}}{k^2+m_D^2} + v^2
\int_{-\infty}^{\infty} \frac{d\omega}{2\pi\omega} \, \left[ k^2
\cos^2{\theta} \, \rho_L+ \omega^2 \sin^2{\theta} \, \rho_T \right]
\right. && \nn \\ \hskip 7cm \times \left. 2
\frac{\sin^2{\left((\omega-kv\cos{\theta})L/(2v)
\right)}}{(\omega-kv\cos{\theta})^2}\right\}_{\rm ind} \ . \ \ \ \ &&
\eeqa

The expression \eq{mastereq2} for the collisional energy loss is
actually ultraviolet divergent. 
The logarithmic 
UV divergence $\sim \int dk/k$ appears in the (leading) asymptotic $L \to \infty$ behaviour of \eq{mastereq2}, 
\beqa
\label{deltainf}
\frac{-\Delta E_{\infty}(L)}{C_R \alpha_s} = v^2 \int
\frac{d^3\vec{k}}{2\pi^2} \int_{-\infty}^{\infty}
\frac{d\omega}{2\pi\omega} \, \left[ k^2 \cos^2{\theta} \, \rho_L+
\omega^2 \sin^2{\theta} \, \rho_T \right]_{\rm ind} \, \left[
\frac{\pi L}{v} \delta(\omega-kv\cos{\theta}) \right] \ . \nn \\ 
\eeqa
In Ref.~\cite{BT} it is stressed that the macroscopic description is meaningful 
only for distant collisions, and a framework which properly includes close collisions is formulated.
Within such an approach the energy loss is UV convergent, but receives a contribution from 
the `hard' domain $k \gg T$. In QCD the running of $\alpha_s$ 
improves the UV convergence, and the stationary rate of energy loss 
$(-dE/dx)_{\infty} = -\Delta E_{\infty}(L)/L$ is expected to be actually dominated
by the `soft' $k \sim m_D$ region \cite{peshierdok}. 
We will come back to this point in the end of section 3.2. 
In the absence of a rigorous treatment with running coupling, we choose to focus on the {\it
difference} between $-\Delta E(L)$ and the standard stationary law \eq{deltainf}, 
\beqa
\label{ddef}
d(L) &=& -\Delta E(L) + \Delta E_{\infty}(L) \\
\label{d}
\frac{d(L)}{C_R \alpha_s} &=& \int \frac{d^3\vec{k}}{2\pi^2} \left\{
\frac{1-\cos{(kL\cos{\theta})}}{k^2+m_D^2} + v^2
\int_{-\infty}^{\infty} \frac{d\omega}{2\pi\omega} \, \left[ k^2
\cos^2{\theta} \, \rho_L+ \omega^2 \sin^2{\theta} \, \rho_T \right]
\right.  \nn \\ && \hskip 2cm \times \left. \left[ 2
\frac{\sin^2{\left((\omega-kv\cos{\theta})L/(2v)
\right)}}{(\omega-kv\cos{\theta})^2} -\frac{\pi L}{v}
\delta(\omega-kv\cos{\theta}) \right] \right\}_{\rm ind} \ , \nn \\
\eeqa 
which turns out to be UV convergent (see Appendix A). 
As discussed in the next section, we also expect the main feature of the function 
$d(L)$ (which we evaluate with fixed coupling), namely its limiting value when $L \to \infty$,  
to be unaffected by the running of $\alpha_s$. 

\subsection{Numerical results and discussion}
\label{sec3.2}

In this section we first discuss the main features of $d(L)$, in particular 
its large $L$ limit, and the domain of validity of our calculation.
(The mathematical properties of $d(L)$ are studied in the Appendices.)  
Consequences on the phenomenology of the collisional loss $-\Delta E(L)$ are then 
presented. 

\begin{figure}[t]
\begin{center}
\epsfig{figure=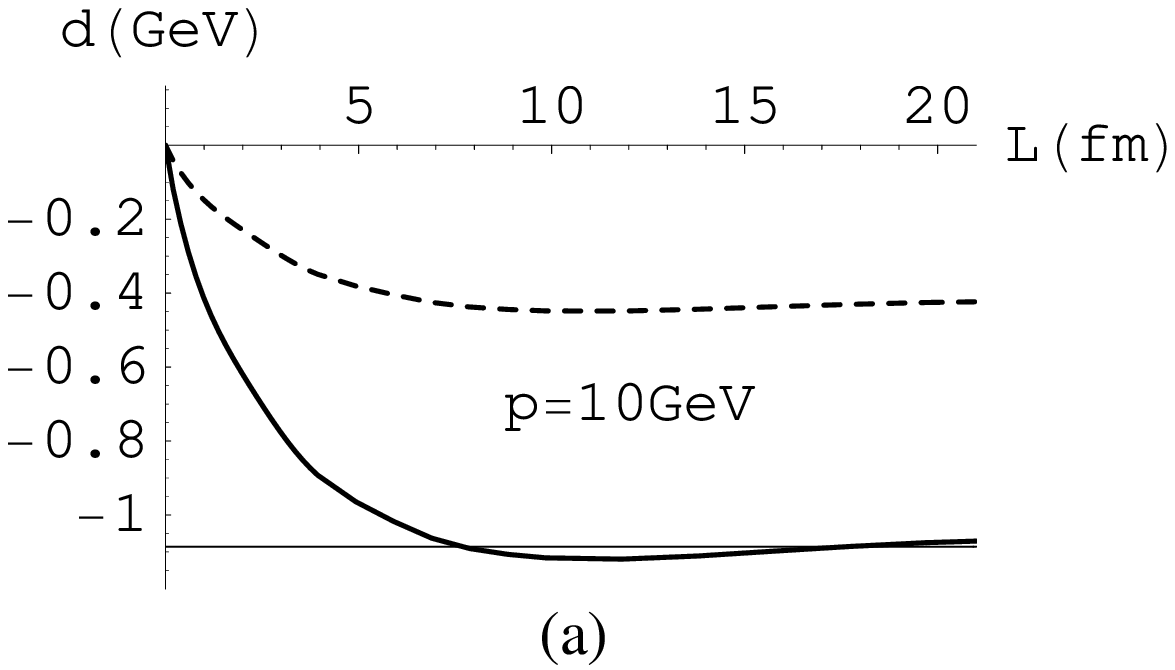,angle=0,width=7cm}
\epsfig{figure=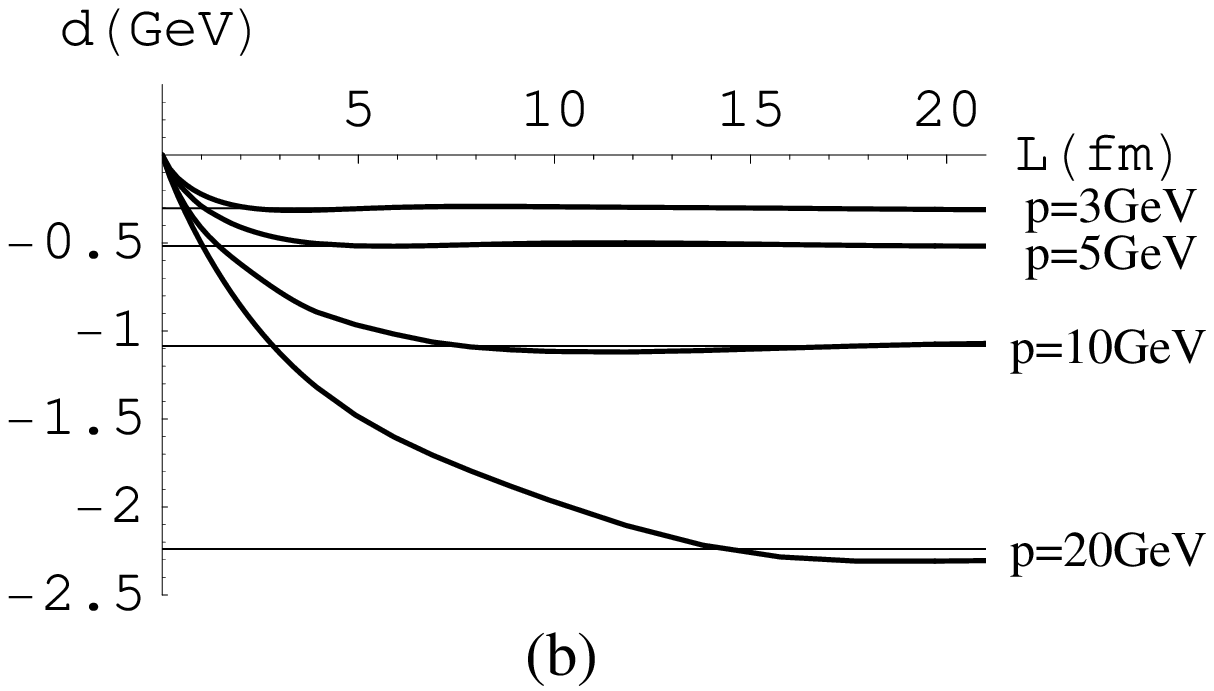,angle=0,width=7cm}
\end{center}
\caption[*]{The function $d(L)$ defined by \eq{ddef} and \eq{d} for a
charm quark ($M=1.5\,{\rm GeV}$) produced in a QGP of temperature
$T=0.25\,{\rm GeV}$, as a function of the distance $L$ travelled in
the plasma. We use $\alpha_s = 0.2$ and $n_f=2$ in the expression of
the Debye mass $m_D^2= 4\pi\alpha_s T^2(1+n_f/6)$, giving $m_D \simeq
0.46\,{\rm GeV}$. (a) $d(L)$ for a quark momentum $p=10\,{\rm GeV}$
(full line). The contribution from initial bremsstrahlung (see section
4 and Appendix B) to $d(L)$ is represented by the dashed line.  (b)
Dependence of $d(L)$ on the charm quark momentum $p$. The thin
straight lines give the values of $d_\infty$ for the corresponding
values of $p$.}
\label{dplot}
\end{figure}

The large $L$ limit of $d(L)$, 
\begin{equation}
d_\infty \equiv \mathop{\rm lim}_{L \to
\infty} d(L) = \mathop{\rm lim}_{L \to \infty} \left[ - \Delta E(L) +
\Delta E_{\infty}(L) \right] \ \ ,
\end{equation}
is calculated exactly in Appendix B (see Eqs.~\eq{al}, \eq{at} and \eq{shiftresult}) 
and can be accurately approximated by  
\begin{equation}
d_\infty\approx -C_R\alpha_s m_D \Bigl(1+\sqrt{2}(\gamma-1)\Bigr) \ \ .
\end{equation}
Our central result is that $d_\infty$ scales as $\gamma = 1/\sqrt{1-v^2}$ when 
$v=p/E \to 1$. The largeness (and negative sign) of $d_\infty$ for large quark energies 
will be the main reason for the important delay of the stationary regime. 
The $L$-dependence of $d(L)$ is presented in Fig.~\ref{dplot} for a
fast charm quark. The function $d(L)$ is found to be negative for all $L$, 
and Fig.~\ref{dplot}b shows the increase (in magnitude) of $d_\infty$ 
with the quark energy, as found analytically. 

The observed values of $d_\infty$ can be translated to an estimate 
of the time scale characteristic of the transitory regime by defining the \emph{retardation time} 
\begin{equation}\label{tret}
t_\mathrm{ret} = d_\infty/(dE/dx)_{\infty}\ \ .
\end{equation}
For a quark with $p=10$~GeV, $-d_\infty\sim 1$~GeV and
$-(dE/dx)_{\infty}\sim 0.1-0.2$~GeV$/$fm yields $t_\mathrm{ret}\sim
5-10$~fm. This number should be considered as a rough estimate, due to the numerous 
approximations used in our theoretical model.
But we stress that the retardation
time may be of the order of several fm for $p=10$~GeV, 
and that it scales with the quark energy for large energies. 

The scaling in $\gamma$ of $d_\infty$ (when $\gamma \to \infty$)
results in a similar scaling of the retardation time, $t_\mathrm{ret}
\sim \gamma /m_D$, and this has the following simple interpretation.
The stationary regime for energy loss sets in when the regions of the
plasma at a transverse distance $\sim 1/m_D$, polarized by the parton
current, start to retroact on the parton. When the parton velocity $v$
is small, this takes a time $t_\mathrm{ret} \sim 1/m_D$. When $v \to
1$, the latter scenario can occur only if the parton is not ahead of
the relevant polarized regions after the time $t_\mathrm{ret}$. At
initial time $t=0$, this requires the parton to polarize a domain
which is ahead of it, at an angle $\theta \sim \sqrt{1-v^2} =1/\gamma
\ll 1$ with respect to the direction of $\vec{v}$. The time $t_\mathrm{ret}$
corresponds to the time necessary to send a signal at a transverse
distance $\sim 1/m_D$ in the direction $\theta$, leading to
$t_\mathrm{ret}\sim 1/(m_D\theta)$ for $\theta\ll 1$, hence
$t_\mathrm{ret} \sim \gamma /m_D$ when $v \to 1$.

We now discuss the domain of validity of our calculation. 
As shown in Appendix B (see \eq{typk}), the typical values of $k$
contributing to $d_\infty$ are $k \sim \morder{\gamma m_D}$. Since
using the HTL gluon spectral densities in \eq{d} a priori requires $k
\ll T$ \cite{pisarski,BI}, our calculation of $d_\infty$ would seem to
be justified, in the perturbative framework $m_D \propto gT \ll T$,
only provided $\gamma$ is not too large, $\gamma \ll 1/g$.  However,
as can be easily inferred from Appendix B, the dominant contribution
to $d_\infty$ arises from $|\omega^2- k^2| \sim m_D^2$ (with $\omega
\simeq k \sim \gamma m_D$), \ie\ from the region where $K=(\omega, k)$
is close to its mass-shell.  It is known \cite{PST,bipp} that in this
region of low virtualities the HTL gluon propagator is a very good
approximation to the exact propagator even in the domain $k \gg T$.
Thus our calculation of $d_\infty$ might be justified for all values
of $\gamma$.

Contrary to the position of the asymptote $d_\infty$, the small $L$
behaviour of $d(L)$ seen on Fig.~\ref{dplot} should not be physically
sound. The small $L$ behaviour of $d(L)$ can be obtained for instance from
\eq{shift2}, and is of the form $d(L \ll m_D^{-1}) \propto \alpha_s^2
T^2 L \log{(m_D L)}$, where the logarithm arises from an integral
$\sim \int_{m_D}^{1/L} dk/k$.  Thus the small $L$ limit is sensitive
to large virtualities $|\omega^2- k^2| \sim k^2 \gg T^2$, where the
HTL approximation becomes invalid. 

In order to qualitatively illustrate how $d(L)$ delays the collisional
energy loss, we have to add the stationary
contribution $-\Delta E_{\infty}(L)$ given in \eq{deltainf}. In the
end of section 3.1 we pointed out that \eq{deltainf} is ill-defined as
it stands, because of a logarithmic UV divergence.
We mentioned that this divergence would be absent, either with a proper treatment of 
close collisions with fixed $\alpha_s$ \cite{BT}, or due to the running of $\alpha_s$
\cite{peshierdok}, both approaches differing by the hardness of $k$ contributing  
to $-(dE/dx)_{\infty}$. This stresses the need for a rigorous treatment with running $\alpha_s$. 
In the absence of such a treatment, and since we need the stationary
loss $-\Delta E_{\infty}(L)$ only for a qualitative illustration of
the retardation effect, we will use the `standard' result derived in Ref.~\cite{BT}, which reads
to logarithmic accuracy (which is sufficient for our purpose):
\beq
\label{TGresult2}
\left( -\frac{dE}{dx} \right)_{\infty} \equiv \frac{-\Delta
E_{\infty}(L)}{L} = \frac{C_R \alpha_s m_D^2}{2} \, \left[ \frac{1}{v}
- \frac{1-v^2}{2v^2}\log{\left( \frac{1+v}{1-v}\right) } \right]
\log{\left(\frac{k_\mathrm{max}}{m_g} \right)}\ , \eeq 
\beq
\label{UVcut}
{\rm where} \hskip .5cm k_\mathrm{max} \equiv {\rm Min \,}\left\{ \frac{ET}{M},
\sqrt{ET} \right\} \ ,  \eeq
and $m_g = m_D/\sqrt{3}$ is the gluon thermal mass.

\begin{figure}[t]
\begin{center}
\epsfig{figure=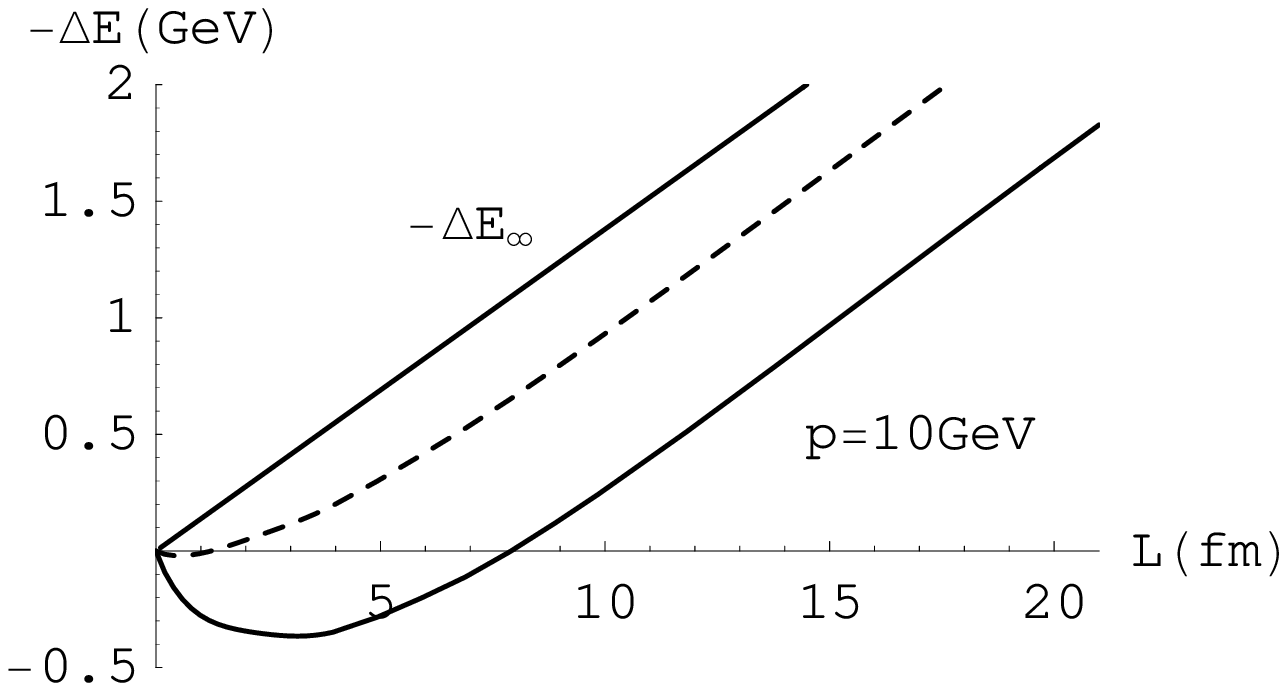,angle=0,width=7cm}
\epsfig{figure=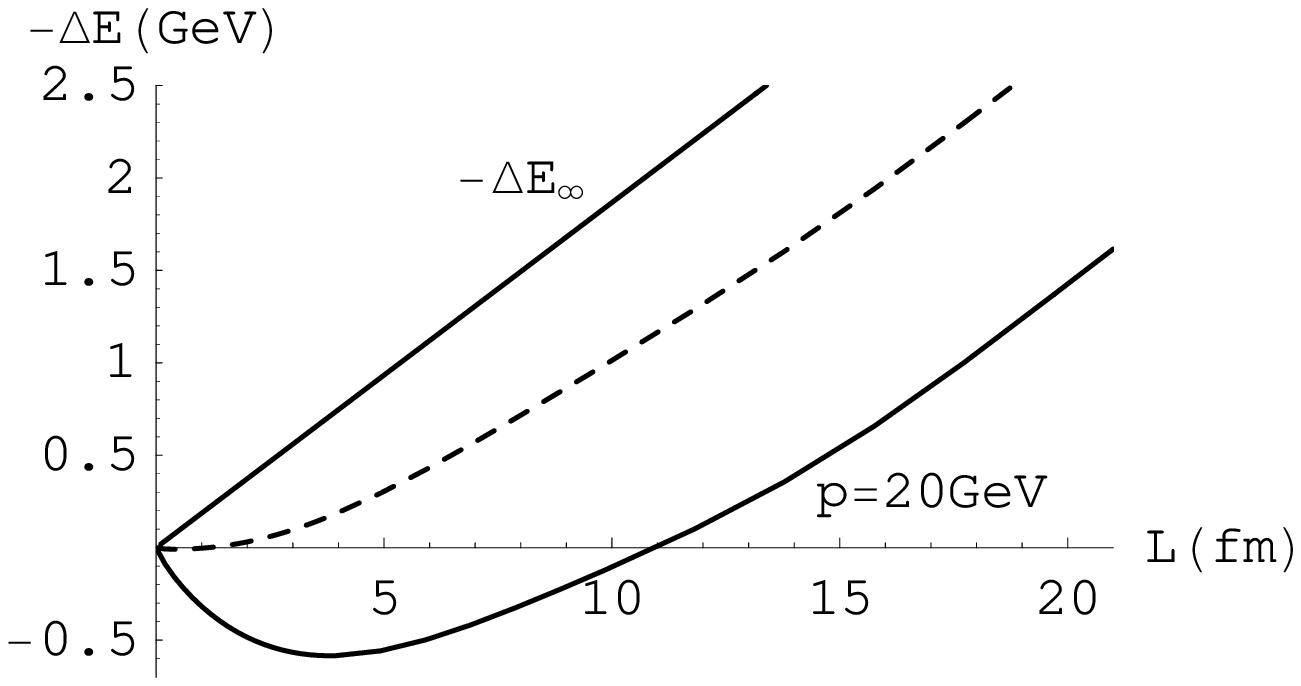,angle=0,width=7cm}
\end{center}
\caption[*]{Collisional energy loss of a charm quark produced in a QGP
as a function of the distance $L$ travelled in the plasma, for $p =
10\,{\rm GeV}$ and $p = 20\,{\rm GeV}$.  The total result (full line)
is compared to the linear law ($-\Delta E_{\infty}$) corresponding to
\eq{TGresult2}, 
and which would arise in Ref.~\cite{BT} by evaluating the slope to
logarithmic accuracy.  As in Fig.~\ref{dplot}a the dashed line singles
out the contribution to $-\Delta E(L)$ from initial bremsstrahlung
(see section 4).}
\label{lossplot}
\end{figure}

Our results for $-\Delta E(L)$ given by \eq{ddef}, \eq{d},
\eq{TGresult2} and \eq{UVcut} are shown in Fig.~\ref{lossplot},
illustrating the main qualitative feature, i.e. the important delay of
the stationary regime. The retardation time \eq{tret}, which
corresponds to the value of $L$ where the asymptote of $-\Delta E(L)$
cuts the horizontal axis, is close to the intersection of the curve
$-\Delta E(L)$ with this axis.

As another consequence of the large negative values of $d(L)$, we
observe on Fig.~\ref{lossplot} that the {\it induced} energy loss
$-\Delta E(L)$ is negative at relatively small $L$, and thus
corresponds to an (induced) energy gain.

Two effects may explain the latter observation, which goes beyond our
initial expectation of a delayed but however positive energy
loss. First, we recall that current conservation requires the
energetic parton to be produced with at least another parton. In the
simple case of a color singlet dipole we considered (see
\eq{genericcurrent}), we expect the dipole to separate more easily in
the medium than in vacuum, due to charge screening. Secondly, the
current created at $t=0$ produces radiation\footnote{The induced
radiated energy is part of $-\Delta E(L)$ as we defined it (see
section 4).}. As discussed in detail in section 4, in the medium the
radiated energy corresponds to the excitation of the QGP plasmon modes
and is reduced compared to vacuum (the {\it induced} radiated energy
is negative), as might have been expected for massive modes. However,
the retardation effect is only {\it partly} due to this difference
between in-medium and vacuum radiation, as seen in Fig.~\ref{dplot}a
and further discussed in section 4. 

We end this section by discussing which features of our calculation might be 
affected by the running of $\alpha_s$. 

First, the standard result \cite{BT} for $-\Delta E_{\infty}(L)$ we have used arises from the
logarithmic interval $m_g \ll k \ll k_\mathrm{max}$ (see \eq{TGresult2} and \eq{UVcut}). 
In QCD, with a {\it running} coupling evaluated at a scale on the order of the gluon 
virtuality, we expect $(-dE/dx)_{\infty}$ to be 
$\propto\int dk \, \alpha_s(k)^2 /k \propto \int dk/(k\,\log^2{k})$,
showing that $(-dE/dx)_{\infty}$ depends negligibly on $E$ when $E \to \infty$ \cite{peshierdok}. 
(In this sense the $\log{k_{\rm max}} \propto \log{E}$ factor in \eq{TGresult2} is an artefact of the 
fixed coupling approximation.)  
Most importantly, $(-dE/dx)_{\infty}$ is actually dominated by the soft
(infrared) region $k \sim m_D$ when $E \gg m_D$ \cite{peshierdok}, and we expect 
the calculation of $(-dE/dx)_{\infty}$ within a macroscopic description and with running 
$\alpha_s$ to be self-consistent.

Secondly, the running of $\alpha_s$ should affect the behaviour of $d(L)$ 
at small $L < 1/m_D$. We discussed previously that since the slope of $d(L)$ at small $L$
is of the form $\sim \alpha_s^2 \int_{m_D}^{1/L} dk/k$, it cannot be consistently derived 
within a macroscopic description. This should not be the case with running coupling, since 
we expect the slope to be rather 
$\sim \int_{m_D}^{1/L} dk/(k \log^2k)$, which is dominated by $k \sim m_D$ and
independent of $L$ when $L\to 0$. Thus the sharp behaviour $d(L \to 0)
\propto L \log{L}$ obtained for fixed $\alpha_s$ should become $d(L
\to 0) \propto - \alpha_s(m_D) T^2 L$ in the case of running
$\alpha_s$. As a consequence, the energy loss $-\Delta E(L) = -\Delta
E_{\infty}(L) + d(L)$ shown in Fig.~\ref{lossplot} should be at most\footnote{In fact, for running 
coupling the slope of $-\Delta E(L)$ is expected to vanish
  when $L\to 0$. This can be seen by expanding the bracket in
  \eq{mastereq} to order $L$. The UV convergence of the integral over $k$ being ensured 
  by the running of the coupling, the integral over
  $\omega$ can be performed by closing the contour in the upper half-plane where there
  is no singularity and identically vanishes.} $\sim \morder{L}$ at small $L$.
Thus, the relatively important induced energy gain at small $L$ seen in
Fig.~\ref{lossplot} might be strongly affected by the running of the coupling. 
In contradistinction, we stress that the large $L$ limit $d_\infty$ of $d(L)$ 
should not be affected by the running of
$\alpha_s$, since $d_\infty$ depends on low virtualities $|\omega^2-k^2| \sim m_D^2$.

\section{Induced radiation}

As already mentioned, and as is well-known, the sudden acceleration of
the energetic parton at time $t=0$ comes along with bremsstrahlung
radiation. The quantity $-\Delta E(L)$ which we called induced {\it
collisional} energy loss actually contains this contribution. In this
section we single out this radiative component from $-\Delta E(L)$ and
emphasize that it plays only a finite part in the retardation effect
studied in section 3.

In order to obtain the contribution $W(L)$ to $-\Delta E(L)$
originating from ra\-dia\-tion, we single out in \eq{r1} the
contribution of the plasmon poles. This is done by using the
expression of the gluon propagators close to their poles \cite{BI},
\ie\ at $\omega^2 \simeq \omega_s^2(k)$, \beqa
\label{gluonresidues}
\Delta_s(\omega,k) \simeq \frac{-z_s(k)}{(\omega +i\eta)^2
-\omega_s^2(k)} = -z_s(k) \left[ {\rm P}\left( \frac{1}{\omega^2
-\omega_s^2(k)}\right) - i\pi {\, \rm sgn}(\omega)\,\delta(\omega^2
-\omega_s^2(k))\right] \ , \nn \\ \eeqa and picking only the
$\delta$-term in the latter. From \eq{r1} we obtain the {\it total}
(\ie, we do not subtract the vacuum contribution from it) in-medium
radiation spectrum: \beqa \frac{d W(L)}{dk \, d\cos{\theta}} =
\frac{C_R \alpha_s}{\pi} \left\{ z_L(k) \frac{k^2}{\omega_L^2(k)}
\cos^2{\theta} \, \frac{\sin^2((\omega_L(k)-kv\cos{\theta})
\,L/(2v))}{(\cos{\theta}-\omega_L(k)/(kv))^2} \right. \hskip 1cm &&
\nn \\ + \left. z_T(k) \sin^2{\theta} \,
\frac{\sin^2((\omega_T(k)-kv\cos{\theta})
\,L/(2v))}{(\cos{\theta}-\omega_T(k)/(kv))^2} \right\} \hskip 1cm &&
\label{indrad} \\ \mathop{\longrightarrow}_{L\to \infty} \frac{C_R
\alpha_s}{2\pi} \left\{ z_L(k) \frac{k^2}{\omega_L^2(k)} \,
\frac{\cos^2{\theta}}{(\cos{\theta}-\omega_L(k)/(kv))^2} + z_T(k) \,
\frac{\sin^2{\theta}}{(\cos{\theta}-\omega_T(k)/(kv))^2} \right\} &&
\label{indradb}
\eeqa where the functions $\omega_{L,T}(k)$ and $z_{L,T}(k)$ can be
found in Ref.~\cite{BI}.

\begin{figure}[t]
\begin{center}
\epsfig{figure=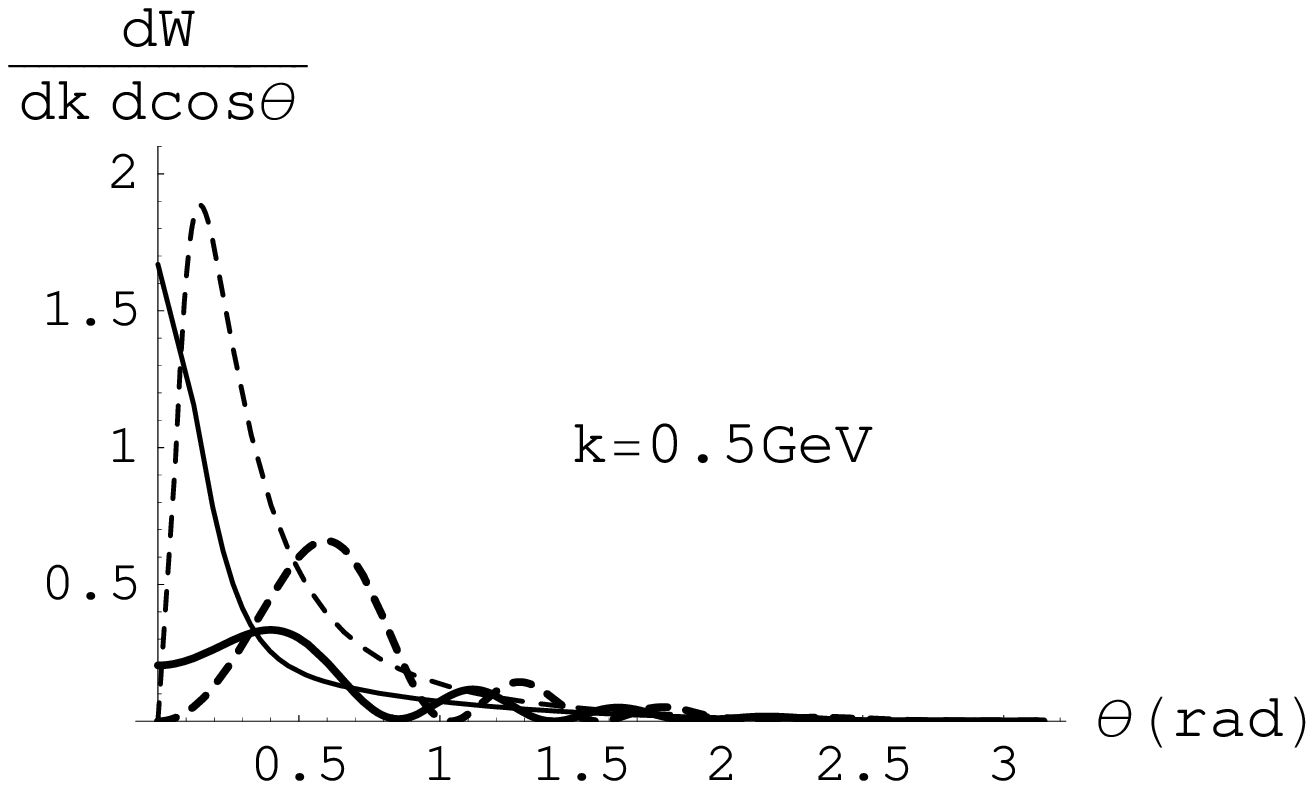,angle=0,width=7cm}
\epsfig{figure=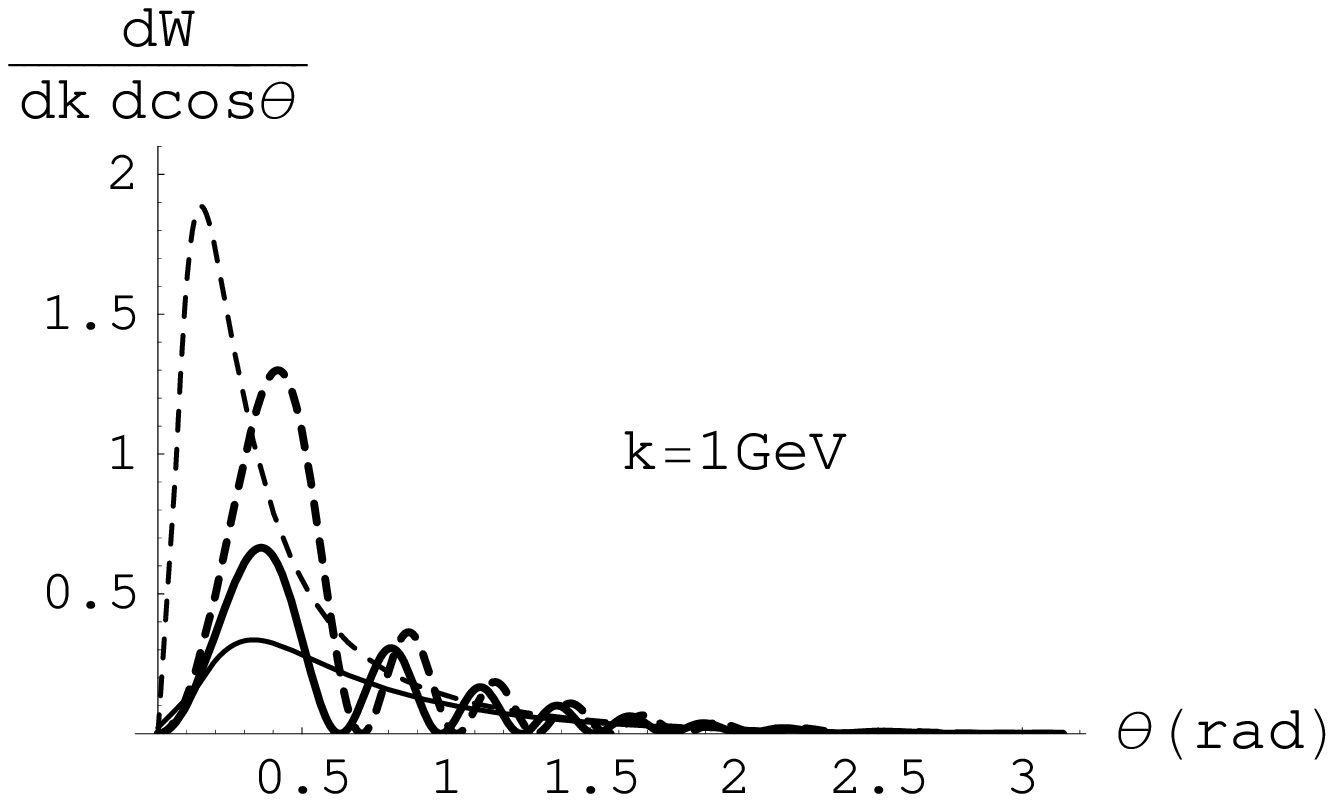,angle=0,width=7cm}
\end{center}
\caption[*]{In-medium angular radiation spectrum (broad full line)
\eq{indrad} for $L= 5\,{\rm fm}$, $p=10\,{\rm GeV}$, $k=0.5 \,{\rm
GeV}$ (left) and $k=1 \,{\rm GeV}$ (right). The thin full line
represents the in-medium spectrum in the $L\to \infty$ limit
\eq{indradb}.  The broad and thin dashed lines represent the same
spectra in the vacuum, given respectively by \eq{indradvac} and
\eq{indradvacb}.}
\label{radplot}
\end{figure}

The differential spectrum in the polar angle $\theta$ is represented
in Fig.~\ref{radplot} for $L= 5\,{\rm fm}$, $p=10\,{\rm GeV}$, and for
two values of $k$, $k=0.5 \,{\rm GeV}$ and $k=1 \,{\rm GeV}$.  The
vacuum contribution is obtained by setting $z_L=0$, $z_T=1$,
$\omega_L(k) = \omega_T(k) = k$ in \eq{indrad}, yielding: \beqa
\left. \frac{d W(L)}{dk \, d\cos{\theta}} \right|_{vac} &&= \frac{C_R
\alpha_s}{\pi} \sin^2{\theta} \, \frac{\sin^2((k-kv\cos{\theta})
\,L/(2v))}{(\cos{\theta}-1/v)^2} \label{indradvac} \\
&&\mathop{\longrightarrow}_{L\to \infty} \frac{C_R \alpha_s}{2\pi}
\frac{v^2 \sin^2{\theta}}{(1-v \cos{\theta})^2} \ . \label{indradvacb}
\eeqa As expected, \eq{indradvacb} corresponds to the vacuum
bremsstrahlung spectrum of a single charge suddenly accelerated at
$t=0$. We remark that the latter spectrum contributes to the radiative
energy loss induced by parton multiple scattering
\cite{bdmps96,Zakharov,GLV}, as the zeroth order in the number of
parton rescatterings (`self-quenching' \cite{GLV}). As already
mentioned, the corresponding in-medium radiation spectrum \eq{indradb}
arises from medium polarization, and thus differs from the spectrum
induced by (at least one) rescattering calculated in
\cite{bdmps96,Zakharov,GLV}.  The (large $L$) in-medium radiation
spectrum \eq{indradb} was previously obtained in Ref.~\cite{DG} within
a diagrammatic approach.

In vacuum the spectrum \eq{indradvacb} is modified by parton cascade
and ha\-dro\-ni\-za\-tion, but the angular pattern \eq{indradvacb} in
principle affects the distribution of final state hadrons. In a QGP of
very large size, this distribution will rather be sensitive to the
large $L$ in-medium spectrum \eq{indradb}. Similarly, the modification
of the finite $L$ diffraction pattern when one goes from vacuum
(spectrum \eq{indradvac}) to the medium (spectrum \eq{indrad}) may be
of some relevance to discuss particle production induced by a fast
parton travelling a distance $L$ in the QGP. When $k$ is large, the
spectrum~\eq{indrad} approaches the vacuum spectrum \eq{indradvac}.
This can already be seen for $k=1 \,{\rm GeV}$ (Fig.~\ref{radplot}
right), although some attenuation is still visible at small $\theta$.
When $k$ decreases (Fig.~\ref{radplot} left), a strong distorsion
shows up as \eq{indrad} becomes sensitive to the plasmon modes of the
QGP.

However the spectrum \eq{indrad} is not directly observable (not
speaking of parton cascade nor hadronization). In a realistic
situation, the parton propagates in the thermal medium for $0 \leq t
\leq L/v$, and in vacuum for $t \geq L/v$. Thus the full picture
should incorporate both stages. In addition, the fast parton produces
{\it transition radiation} when going through the discontinuity
between the thermal medium ($\epsilon_{L,T}\neq 1$) and the vacuum
($\epsilon_{L,T}= 1$).  This transition radiation must also be taken
into account in order to obtain the complete radiation spectrum. We
postpone to a future work the study of transition radiation in our
context, as well as its influence on the overall energy loss. In
particular, it is too early to possibly relate the angular spectrum
\eq{indrad} (see Fig.~\ref{radplot}) to the humpbacked azimuthal
distribution of particles produced back to the leading jet observed at
RHIC \cite{rhicaz}.  However, the angular radiation spectrum
\eq{indrad} shown in Fig.~\ref{radplot} suggests that (at least some
component of) the particle yield in heavy ion collisions should
exhibit a diffraction pattern arising from the {\it finite length} of
the medium along which collisional energy loss occurs.
 
Finally, we stress that in our calculation the radiated energy
(obtained by integrating \eq{indrad} over the radiated gluon energy
and emission angle) is smaller in the medium than in vacuum. This
feature could be foreseen by comparing the in-medium and vacuum
dif\-fe\-ren\-tial spectra in Fig.~\ref{radplot}. Thus the
contribution of induced radiation to what we defined as the induced
collisional energy loss $-\Delta E(L)$ is negative. Numerically, we
find that the bremsstrahlung contribution always accounts for less
than $50 \%$ of the difference $d(L)$ between $-\Delta E(L)$ and the
asymptotic result $-\Delta E_{\infty}$ (see Figs.~\ref{dplot}a and
\ref{lossplot}). In the end of Appendix B we show that the relative
contribution from bremsstrahlung to $d(L)$ reaches exactly $1/2$ when
$L\to \infty$ and in the ultrarelativistic limit $v \to 1$. This shows
that the retardation effect studied in section 3 cannot be solely
attributed to the radiative component of $-\Delta E(L)$.

\section{Conclusion}

We have studied the collisional energy loss $-\Delta E(L)$, in the
fixed coupling approximation, of an energetic parton travelling the
distance $L$ in a QGP, and initially produced (at $t=0$) in the
medium. Compared to previous estimates which assumed the parton to be
produced at $t=-\infty$, $-\Delta E(L)$ is strongly suppressed up to
$L \sim t_\mathrm{ret}$, where the retardation time $t_\mathrm{ret}$
scales with the parton momentum $p$ at large $p$.  For $p=10 \,{\rm
GeV}$ we roughly estimated $t_\mathrm{ret} \sim 5-10\,{\rm fm}$.
We stress that this estimate is only indicative due to the various 
approximations used in our theoretical model, in particular the small coupling 
limit $g \ll 1$. Also, the running of $\alpha_s$ is expected to 
affect both the small $L$ behaviour of $-\Delta E(L)$, and the 
asymptotic stationary rate $(-dE/dx)_{\infty}$, as discussed in the end of section 
3.2. A rigorous treatment with 
running $\alpha_s$ would be needed to obtain a better quantitative estimate
of $t_\mathrm{ret}$.

We believe our results could be relevant to jet quenching phenomenology, 
since it has recently been argued
\cite{MustafaThoma,Mustafa} that collisional energy loss should be
reconsidered as an important source of energy loss. In addition to the
suppression of $-\Delta E(L)$, we find that the asymptotic linear
behaviour of $-\Delta E(L)$ is delayed to about $L \sim
t_\mathrm{ret}$. In particular, using a stationary energy loss rate
$-dE/dx$ makes sense only for quite large values of $L >
t_\mathrm{ret}$.
 
The suppression and retardation of $-\Delta E(L)$ are encoded in the
difference $d(L) = -\Delta E(L) +\Delta E_{\infty}(L)$ between
$-\Delta E(L)$ and the `standard' stationary result $-\Delta
E_{\infty}(L) \propto L$. As we have shown, $d(L)$ is a well-defined
(UV convergent) quantity, and our main result - the {\it largeness} of the
retardation time - is independent of the uncertainties on $(-dE/dx)_{\infty}$
and on the precise shape of $-\Delta E(L)$ at small $L$. 
The main reason for the large magnitude of the
retardation time is the scaling $d_\infty \propto - \gamma m_D$ when
$\gamma = E/M \to\infty$, where $d_\infty$ is the large $L$ limit of
$d(L)$.  We showed that the (induced) bremsstrahlung arising from the
hard parton being suddenly accelerated at $t=0$, formally included in
$d(L)$, contributes only partly to $d_\infty$.  Thus the retardation
effect cannot be solely attributed to initial radiation. The scaling
in $\gamma$ of $d_\infty$ (when $\gamma \to \infty$) results in a
similar scaling of the retardation time, $t_\mathrm{ret} \sim \gamma
/m_D$, and a physical interpretation of this fact is given in section 3.2.

Finally, we mention that in order to explain the observed dependence
\cite{dEnterria} in azimuthal angle $\phi$ (with respect to the
reaction plane in a heavy ion $AA$ collision) of the nuclear
modification factor $R_{AA}$, taking into account the geometry of the
collision is not sufficient \cite{Pantuev}. A length scale $L \simeq
2\,{\rm fm}$ has to be introduced, below which the parent parton of
the high $p_T$ jet or hadron is assumed to be insensitive to energy
loss. In Ref.~\cite{Pantuev} this parameter is interpreted as the
formation time of the plasma, but it is also stressed there that
$2\,{\rm fm}$ is quite large compared to the values $\sim 0.2\,{\rm
fm}$ usually taken for the plasma formation time.  Our study suggests
that this parameter might instead hint to the possibility of a {\it
negative} loss $-\Delta E_{coll}(L)$ before the stationary regime,
partially compensating the radiative energy loss $-\Delta E_{rad}(L)$
induced by rescattering of the energetic parton.

\vskip 1cm \acknowledgments We would like to warmly thank D.~Schiff,
Y.~L.~Dokshitzer and A.~Peshier for very instructive and helpful
discussions.  We are also grateful to J.~Aichelin for stimulating
exchanges during this work.

\bigskip
\centerline{APPENDIX}

\appendix

\section{Ultraviolet convergence of the function $d(L)$}

Here we show that $d(L)$ given by \eq{d} is ultraviolet
convergent. For this purpose, it is sufficient to prove that the
angular integral \beq
\label{Idef}
I = \int_{-1}^1 d \cos{\theta} \, f(\cos{\theta}) \left[ 2
\frac{\sin^2{\left(\frac{kL}{2v}(x-v\cos{\theta})
\right)}}{k^2(x-v\cos{\theta})^2} -\frac{\pi L}{k v^2} \delta\left(
\frac{x}{v}- \cos{\theta} \right) \right] \ \ , \eeq where
$x=\omega/k$ and $f(\cos{\theta}) = \cos^2{\theta}$ or
$\sin^2{\theta}$, is of order $\sim \morder{1/k^2}$ when $k \to
\infty$. Indeed, the second term of \eq{Idef} is of order $\sim
\morder{1/k}$, leading to the logarithmic UV divergence of $-\Delta
E_{\infty}(L)$ (see \eq{deltainf}). 
With the change of variable \beq
\label{cv}
u = \alpha \, (x - v \cos{\theta}) \ \ \ ; \ \ \ \alpha \equiv \frac{k
L}{2 v} \eeq we obtain \beqa I = \frac{L}{k v^2} \left\{
\int_{\alpha(x - v)}^{\alpha(x +v)} du \frac{\sin^2{u}}{u^2}
f\left(\frac{x}{v}-\frac{u}{\alpha v} \right) - \pi \theta(v -|x|)
f\left(\frac{x}{v} \right) \right\} \equiv I_1 + I_2 \ \ , \eeqa where
we define \beqa
\label{i1}
I_1 &=& \frac{L}{k v^2} f\left(\frac{x}{v} \right)
\left[\int_{\alpha(x - v)}^{\alpha(x +v)} du \frac{\sin^2{u}}{u^2} -
  \pi \theta(v -|x|) \right] \ \ , \\
\label{i2}
I_2 &=& \frac{L}{k v^2} \int_{\alpha(x - v)}^{\alpha(x +v)} du
\frac{\sin^2{u}}{u^2} \left( f\left(\frac{x}{v}-\frac{u}{\alpha v}
\right) - f\left(\frac{x}{v} \right) \right) \ \ .  \eeqa

The behaviour of $I_1$ when $\alpha = k L/(2 v) \to \infty$ is easily
found by treating separately the two cases $|x| > v$ and $|x| < v$. In
the first case we can replace $\sin^2{u} \to 1/2$ in the integrand and
we find \beq
\label{i1limit}
I_1 \mathop{\sim}_{\ \ \alpha \to \infty\ \ } \frac{L}{k v^2}
f\left(\frac{x}{v} \right) \frac{v}{\alpha} \frac{1}{x^2-v^2} =
\frac{2}{k^2} f\left(\frac{x}{v} \right) \frac{1}{x^2-v^2} \ \ , \eeq
which is $\sim \morder{1/k^2}$. In the second case, $|x| < v$, we
write \beq \int_{\alpha(x - v)}^{\alpha(x +v)} du
\frac{\sin^2{u}}{u^2} - \pi = - \int_{\alpha(x + v)}^{\infty} du
\frac{\sin^2{u}}{u^2} - \int^{\alpha(x - v)}_{-\infty} du
\frac{\sin^2{u}}{u^2} \ \ , \eeq where the replacement $\sin^2{u} \to
1/2$ can be made in the r.h.s. when $\alpha \to \infty$, leading again
to \eq{i1limit}. Note that when $\alpha \to \infty$, the expression
\eq{i1} is thus equivalent to \beq I_1 \mathop{\ \ \sim\ \ }_{\alpha
\to \infty} \frac{L}{2 k v^2} f\left(\frac{x}{v} \right)
\int_{\alpha(x - v)}^{\alpha(x +v)} du \, {\rm P} \left(\frac{1}{u^2}
\right) \ \ .  \eeq

In the $I_2$ integral defined in \eq{i2}, the contribution from $u \ll
\alpha$ is negligible, showing that typically $u \sim \alpha$. Thus we
can replace $\sin^2{u} \to 1/2$, yielding: \beq I_2 \mathop{\ \ \sim\
\ }_{\alpha \to \infty} \frac{L}{2 k v^2} \int_{\alpha(x -
v)}^{\alpha(x +v)} du \, {\rm P} \left(\frac{1}{u^2} \right) \left(
f\left(\frac{x}{v}-\frac{u}{\alpha v} \right) - f\left(\frac{x}{v}
\right) \right) \ \ .  \eeq

We infer from the above that the behaviour of \eq{Idef} when $\alpha =
k L/(2 v) \to \infty$ is obtained by the following replacement: \beq
\label{formalreplace}
\left[ 2 \frac{\sin^2{\left(\frac{kL}{2v}(x-v\cos{\theta})
      \right)}}{k^2(x-v\cos{\theta})^2} -\frac{\pi L}{k v^2}
  \delta\left( \frac{x}{v}- \cos{\theta} \right) \right] \mathop{\ \
  \longrightarrow\ \ }_{\alpha \to \infty} {\rm P} \left( \frac{1}
     {k^2(x-v\cos{\theta})^2} \right) \ \ .  \eeq The r.h.s. is $\sim
     \morder{1/k^2}$, to be compared to $\morder{L/k}$ (second term of
     \eq{Idef}).  This completes our proof of the UV convergence of
     the function $d(L)$.

\section{Large $L$ limit of $d(L)$}

Here we evaluate the limiting value of $d(L)$ (defined by \eq{ddef})
when $L \to \infty$, \beq d_\infty \equiv \mathop{\rm lim}_{L \to
\infty} d(L) = \mathop{\rm lim}_{L \to \infty} \left[ - \Delta E(L) +
\Delta E_{\infty}(L) \right] \ \ .  \eeq

\subsection{Total result for $d_\infty$}

Using \eq{r1} (with the bracket expressed as in \eq{bracket}) we
obtain: \beqa
\label{shift}
\frac{d(L)}{C_R \alpha_s} &=& i v^2 \int \frac{d^3\vec{k}}{4\pi^3}
\int_{-\infty}^{\infty} \frac{d\omega}{\omega} \, \left[ k^2
\cos^2{\theta} \, \Delta_L(\omega, k)+ \omega^2 \sin^2{\theta} \,
\Delta_T(\omega, k) \right]_{\rm ind} \nn \\ &\times& \left\{ \pi \,
\delta(K.V) L/v - 2\,\frac{\sin^2(K.V \,L/(2v))}{(K.V)^2} -
i\,\frac{\sin(K.V \,L/v)}{K.V} {\rm P}\left(\frac{1}{K.V}\right)
\right\} \ .  \eeqa We now use the fact that if we replace $L$ by $-L$
in \eq{r1}, the integral over $\omega$, performed by closing the
integration contour in the {\it upper} half-plane, vanishes
identically because the singularities in $\omega$ all lie below the
real axis: \beqa
\label{zero}
0 &=& i v^2 \int \frac{d^3\vec{k}}{4\pi^3} \int_{-\infty}^{\infty}
\frac{d\omega}{\omega} \, \left[ k^2 \cos^2{\theta} \,
\Delta_L(\omega, k)+ \omega^2 \sin^2{\theta} \, \Delta_T(\omega, k)
\right]_{\rm ind} \nn \\ &\times& \left\{ - \pi \, \delta(K.V) L/v +
2\,\frac{\sin^2(K.V \,L/(2v))}{(K.V)^2} - i\,\frac{\sin(K.V
\,L/v)}{K.V} {\rm P}\left(\frac{1}{K.V}\right) \right\} \ .  \eeqa
Adding \eq{zero} to \eq{shift} we get: \beqa
\label{shift2}
\frac{d(L)}{C_R \alpha_s} &=& i v^2 \int \frac{d^3\vec{k}}{4\pi^3}
\int_{-\infty}^{\infty} \frac{d\omega}{\omega} \, \left[ k^2
\cos^2{\theta} \, \Delta_L(\omega, k)+ \omega^2 \sin^2{\theta} \,
\Delta_T(\omega, k) \right]_{\rm ind} \nn \\ && \hskip 3cm \times
\left\{- 2 i\,\frac{\sin(K.V \,L/v)}{K.V} {\rm
P}\left(\frac{1}{K.V}\right) \right\} \ .  \eeqa Using the first of
the identities \eq{delta}, and then $\delta(x) \, {\rm P}(1/x) = -
\delta'(x)$, we obtain in the $L\to \infty$ limit: \beq
\label{shift3}
\frac{d_\infty}{C_R \alpha_s} = - \pi v^2 \re \int
\frac{d^3\vec{k}}{2\pi^3} \int_{-\infty}^{\infty}
\frac{d\omega}{\omega} \, \left[ k^2 \cos^2{\theta} \,
\Delta_L(\omega, k)+ \omega^2 \sin^2{\theta} \, \Delta_T(\omega, k)
\right]_{\rm ind} \, \delta'(K.V) \ .  \eeq In $\delta'(K.V)$ we trade
the derivative $\partial / \partial \omega$ for $\partial / \partial
v$: \beq
\label{shift4}
\frac{d_\infty}{C_R \alpha_s} = \pi v^2 \frac{\partial}{\partial v}
\frac{1}{v} \re \int \frac{d^3\vec{k}}{2\pi^3} \, \left[ \Delta_L(kv
\cos{\theta}, k)+ v^2 \sin^2{\theta} \, \Delta_T(kv \cos{\theta}, k)
\right]_{\rm ind} \ .  \eeq Using \eq{prop} we have \beqa \left[
\Delta_L(k x,k) \right]_{\rm ind} &=&
\frac{\Pi_L(x)}{k^2\left[k^2+\Pi_L(x)\right]} \\ \left[ \Delta_T(k
x,k) \right]_{\rm ind} &=& \frac{-
\Pi_T(x)}{k^2(x^2-1)\left[k^2(x^2-1)-\Pi_T(x)\right]} \ \ , \eeqa and
denoting $x=v \cos{\theta}$ the equation \eq{shift4} becomes: \beq
\label{shift5}
\frac{d_\infty}{C_R \alpha_s} = v^2 \frac{\partial}{\partial v}
\frac{1}{v^2} \int_{-v}^v dx \int_0^\infty \frac{dk}{\pi} \re \left\{
\frac{\Pi_L(x)}{k^2+\Pi_L(x)} -\frac{v^2-x^2}{1-x^2}
\frac{\Pi_T(x)}{k^2(1-x^2)+\Pi_T(x)} \right\} \ \ .  \eeq We finally
perform the integral over $k$, \beq
\label{finalshift}
\frac{d_\infty}{C_R \alpha_s} = m_D v^2 \frac{\partial}{\partial v}
\frac{1}{v^2} \int_0^v dx \re \left\{ \sqrt{\hat{\Pi}_L(x)} -
\frac{v^2-x^2}{(1-x^2)^{3/2}} \sqrt{\hat{\Pi}_T(x)} \right\} \ \ ,
\eeq where \beqa && \hat{\Pi}_L(x) \equiv \Pi_L(x)/m_D^2 = 1 -
\frac{x}{2} \log{\left( \frac{x+1}{x-1} \right)} \nn \\ &&
\hat{\Pi}_T(x) \equiv \Pi_T(x)/m_D^2 = \frac{x^2}{2} \left[ 1 -
\frac{x^2-1}{2x} \log{\left( \frac{x+1}{x-1} \right)} \right] \ \ .
\eeqa

The longitudinal contribution to \eq{finalshift} reads: \beqa
\label{longshiftresult}
d_{\infty \, L} = - C_R \alpha_s m_D \, A_L(\gamma) \hskip 5cm && \\
\nn \\ 
\label{al} 
A_L(\gamma) = \frac{2}{v} \int_0^v dx \re \sqrt{1-\frac{x}{2}
\log{\left| \frac{x+1}{x-1} \right|} +\frac{i\pi x}{2}} - \re
\sqrt{1-\frac{v}{2} \log{\left| \frac{v+1}{v-1} \right|} +\frac{i\pi
v}{2}} \ . \ \ \ \ \ &&
\label{explicitlongshift}
\eeqa We check numerically that $A_L(\gamma)$ is a smooth increasing
function of the parton Lorentz factor $\gamma = 1/\sqrt{1-v^2}$,
increasing very slowly above $\gamma=10$, and saturating when $\gamma
\to \infty$: $A_L(1)=1$, $A_L(10) \simeq 1.3$, $A_L(1000) \simeq 1.5$,
$A_L(\infty) \simeq 1.814$.

For the transverse contribution to \eq{finalshift} we find: \beqa
\label{transshiftresult}
d_{\infty \, T} = - C_R \alpha_s m_D \, A_T(\gamma) \hskip 5cm &&\\
\nn \\ 
\label{at}
A_T(\gamma) = \frac{2}{v} \int_0^v dx \frac{x^2}{(1-x^2)^{3/2}}
\re \sqrt{\frac{x^2}{2}\left[1-\frac{x^2-1}{2x} \log{\left|
\frac{x+1}{x-1} \right|} \right] +\frac{i\pi x (x^2-1)}{4}} \ . \ \ \
\ &&
\label{explicittransshift}
\eeqa When $v\to 1$, a singularity at $x \to 1$ appears in the
integrand of \eq{explicittransshift}, and we easily derive the
asymptotic behaviour $A_T(\gamma) \simeq \sqrt{2} \,\gamma$ when
$\gamma \to \infty$, giving: \beq
\label{transshiftresultas}
d_{\infty \, T} \mathop{\simeq}_{v\to 1} - \sqrt{2} \, C_R \alpha_s
m_D \, \gamma \ \ .  \eeq Numerically, $A_T(\gamma)$ (which satisfies
$A_T(1)=0$) is quite well approximated by the linear form
$\sqrt{2}(\gamma -1)$ for all values of $\gamma \geq 2$, with an
accuracy of $25 \%$ for $\gamma = 2$ and improving for increasing
$\gamma$.

\begin{figure}[t]
\begin{center}
\epsfig{figure=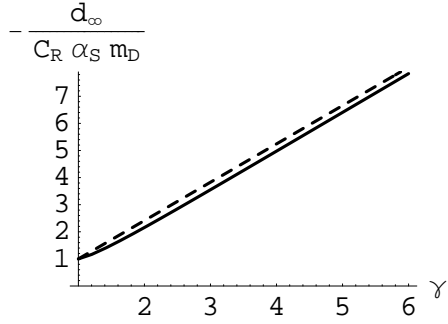,angle=0,width=7cm}
\end{center}
\caption[*]{The exact function $-d_\infty/(C_R \alpha_s m_D)$ (full line) and its 
approximation $1 + \sqrt{2} (\gamma -1)$ (dashed line), as a function of 
$\gamma = E/M$, see \eq{shiftresult}.}
\label{deltashiftfig}
\end{figure}

Adding \eq{longshiftresult} and \eq{transshiftresult} we obtain 
\beq
\label{shiftresult}
d_\infty = - C_R \alpha_s m_D \left( A_L(\gamma) + A_T(\gamma) \right) \simeq - C_R \alpha_s m_D 
\left( 1 + \sqrt{2} (\gamma -1)  \right) \ , 
\eeq
where the latter approximation has the correct limits at $\gamma \to
1$ and $\gamma \to \infty$, and can be checked numerically to be
accurate to better than $12 \%$ for all values of $\gamma$ (see
Fig.~\ref{deltashiftfig}).

It is instructive to find what are the typical values of $k$ and
$\theta$ which contribute to the leading behaviour
\eq{transshiftresultas} of $d_\infty$. In \eq{shift5} the transverse
contribution is dominated by $x = v \cos{\theta} \to 1$ when $v\to 1$,
\ie\ $\theta \ll 1$. Using $\Pi_T(x\to 1) = m_D^2/2$, the leading part
of \eq{shift5} can be written as \beq
\label{typk}
d_{\infty} \mathop{\ \ \simeq\ \ }_{v\to 1} - C_R \alpha_s \int_0^1
\frac{d\theta^2}{\theta^2+1/\gamma^2} \int_0^\infty \frac{dk}{\pi}
\frac{m_D^2}{k^2 \left( \theta^2+1/\gamma^2 \right) + m_D^2/2} \ \ .
\eeq The latter is easily checked to arise from $\theta \sim 1/\gamma$
and $k \sim \gamma m_D$ (as well as to yield the result
\eq{transshiftresultas}).

\subsection{Contribution from radiation to $d_\infty$, in the $v\to 1$
  limit}

In order to see how the bremsstrahlung induced by the initial
acceleration of the parton affects the retardation effect, we single
out in $d_\infty$, given by \eq{shiftresult}, the contribution from
radiation. This contribution $W_\infty$ is obtained by subtracting
\eq{indradvacb} from \eq{indradb}, and then integrating over $k$ and
$\theta$. Here we concentrate on the transverse contribution
$W_{\infty\, T}$ which turns out to be dominant in the $v\to 1$ limit,
\beq
\label{Winf1}
\frac{W_{\infty\, T}}{C_R \alpha_s} = \int_0^\infty \frac{dk}{2\pi}
\int_{-1}^1 d\cos{\theta} \left\{
\frac{z_T(k)\,\sin^2{\theta}}{(\cos{\theta}-\omega_T(k)/(kv))^2} -
\frac{\sin^2{\theta}}{(\cos{\theta}-1/v)^2} \right\} \ \ .  \eeq When
$v\to 1$, the above integral is dominated by the domain $\theta \ll
1$, $k \gg m_D$.  For $k \gg m_D$ we have $\omega_T(k) \simeq k +
m_\infty^2/(2k)$, where $m_\infty = m_D/\sqrt{2}$ is the asymptotic
gluon thermal mass, and $z_T(k) \simeq 1$ \cite{BI} can also be
consistently used in \eq{Winf1}.  Approximating the integrand in
\eq{Winf1} we obtain \beq
\label{Winf2}
\frac{W_{\infty\, T}}{C_R \alpha_s} \mathop{\ \ \simeq \ \ }_{v\to 1}
- \frac{2 m_\infty^2}{\pi} \int_0^\infty \frac{dk}{k^2} \int_0^1
d\theta^2\,\theta^2 \frac{\theta^2+\frac{1}{\gamma^2}
+\frac{m_\infty^2}{2k^2}}{\left(
\theta^2+\frac{1}{\gamma^2}\right)^2\left(
\theta^2+\frac{1}{\gamma^2}+\frac{m_\infty^2}{k^2} \right)^2} \ \ .
\eeq In the ultrarelativistic $\gamma \to \infty$ limit, the typical
values of $k$ and $\theta$ in the latter integral are $k \sim \gamma
m_\infty$ and $\theta \sim 1/\gamma$. The calculation is now
straightforward and we obtain for the leading term: \beq
\label{Winf3}
W_{\infty\, T} \mathop{\ \ \simeq \ \ }_{v\to 1} - C_R \alpha_s m_D
\gamma /\sqrt{2} \ \ .  \eeq This is exactly half of the full result
for $d_{\infty}$, see \eq{transshiftresultas} and \eq{shiftresult}.

For completeness we quote the result for the longitudinal contribution
to $W_{\infty}$, \beq W_{\infty\, L} \mathop{\ \ \simeq \ \ }_{v\to 1}
\frac{2 C_R \alpha_s m_D}{3\pi} \left( \log \gamma \right)^{3/2} \ \ ,
\eeq which is indeed subleading compared to \eq{Winf3} when $v\to 1$.

\end{document}